\documentclass[sigconf]{acmart}

\usepackage{booktabs} 

\setcopyright{none}
\usepackage[linesnumbered,ruled]{algorithm2e}
\usepackage{graphicx}
\usepackage{caption}
\usepackage{caption}
\usepackage{natbib}
\usepackage{amsmath}
\usepackage{soul}
\usepackage{url}
\usepackage{enumerate}
\usepackage[utf8]{inputenc}
\usepackage{mathtools}
\usepackage{subcaption} 
\usepackage{cancel}
\usepackage{comment}
\usepackage{ctable}
\usepackage{mathtools}
\usepackage{balance}
\usepackage{threeparttable}  
\usepackage{booktabs}
\usepackage{multirow}
\usepackage{bm}
\usepackage{enumitem}
\usepackage{amsthm}
\usepackage{xcolor,colortbl}
\theoremstyle{definition}   
\newtheorem{Definition}{Definition}

\newtheorem{Example}{Example}
\newcommand{\todo}[1]{\textcolor{red}{ToDo: #1}}
\newcommand{\specialcell}[2][c]{%
	\begin{tabular}[#1]{@{}c@{}}#2\end{tabular}}
\newcommand{\xhdr}[1]{\vspace{2mm}{\noindent\bfseries #1}.}
\usepackage{pifont}
\newcommand{\cmark}{\text{\ding{51}}}

\usepackage{geometry}
\renewcommand\footnotetextcopyrightpermission[1]{}\renewcommand\footnotetextcopyrightpermission[1]{}

\begin{document}
\title[SHGNN: Structure-Aware Heterogeneous Graph Neural Network]{SHGNN: Structure-Aware Heterogeneous Graph Neural Network}

\author{Wentao Xu$^{1}$*, Yingce Xia$^{2}$, Weiqing Liu$^{2}$, Jiang Bian$^{2}$, Jian Yin$^{1}$, Tie-Yan Liu$^{2}$}
\thanks{*Work done while Wentao Xu was an intern at Microsoft Research}
\affiliation{
	\institution{
		$^1$Sun Yat-sen University\\
		$^2$Microsoft Research Asia\\
	}
	\country{}
}
\email{{xuwt6@mail2,issjyin@mail}.sysu.edu.cn}
\email{{yingce.xia, weiqing.liu, jiang.bian, tyliu}@microsoft.com}
\renewcommand{\authors}{Wentao Xu, Yingce Xia, Weiqing Liu,  Jiang Bian, Jian Yin, Tie-Yan Liu}
\renewcommand{\shortauthors}{Xu et al.}

\begin{abstract}
Many real-world graphs (networks) are heterogeneous with different types of nodes and edges. Heterogeneous graph embedding, aiming at learning the low-dimensional node representations of a heterogeneous graph, is vital for various downstream applications. Many meta-path based embedding methods have been proposed to learn the semantic information of heterogeneous graphs in recent years. However, most of the existing techniques overlook the graph structure information when learning the heterogeneous graph embeddings. This paper proposes a novel Structure-Aware Heterogeneous Graph Neural Network (SHGNN) to address the above limitations. In detail, we first utilize a feature propagation module to capture the local structure information of intermediate nodes in the meta-path. Next, we use a tree-attention aggregator to incorporate the graph structure information into the aggregation module on the meta-path. Finally, we leverage a meta-path aggregator to fuse the information aggregated from different meta-paths. 
We conducted experiments on node classification and clustering tasks and achieved state-of-the-art results on the benchmark datasets, which shows the effectiveness of our proposed method.
\end{abstract}



\maketitle
\section{Introduction}
\label{sec:intro}
Lots of real-world data naturally exists in the graph structure, such as the social networks~\cite{DBLP:conf/kdd/WangC016, DBLP:conf/nips/HamiltonYL17}, the traffic networks~\cite{li2018diffusion, DBLP:conf/uai/ZhangSXMKY18}, the citation networks~\cite{NIPS2016_6212, DBLP:conf/iclr/KipfW17, DBLP:conf/nips/HamiltonYL17}, the physical systems~\cite{NIPS2016_6418,NIPS2017_7231}, and the knowledge graphs~\cite{bordes2013translating,xu-etal-2020-seek}.
The underlying data structure of graph data is non-Euclidean. We cannot directly use the algorithms designed for data in the Euclidean space. 
Learning good embeddings/representations of graph structure data benefits many downstream applications, such as the recommendation systems~\cite{fan2019metapath,zhao2019intentgc,liu2020heterogeneous,jin2020efficient} and the question answering~\cite{teney2017graph,huang2019knowledge,lv2020commonsense,zhao2020complex}.
Researchers had conducted extensive study on graph embeddings, ranging from classic machine learning methods (e.g., LINE~\cite{Tang:2015:LLI:2736277.2741093}, DeepWalk~\cite{Perozzi:2014:DOL:2623330.2623732}, node2vec~\cite{Grover:2016:NSF:2939672.2939754}, TADW~\cite{Yang:2015:NRL:2832415.2832542}, and Struc2vec~\cite{ribeiro2017struc2vec}) to deep learning based methods, i.e., graph neural networks (briefly, GNN). Typical GNN networks include
ChebNet~\cite{Defferrard:2016:CNN:3157382.3157527}, GCN~\cite{DBLP:conf/iclr/KipfW17}, GraphSAGE~\cite{DBLP:conf/nips/HamiltonYL17}, GAT~\cite{DBLP:conf/iclr/VelickovicCCRLB18}, GGNN~\cite{DBLP:journals/corr/LiTBZ15} and GaAN~\cite{DBLP:conf/uai/ZhangSXMKY18}.

\begin{figure}[t]
	\centering
	\begin{subfigure}[b]{0.22\textwidth}
		\centering
		\includegraphics[width=\textwidth]{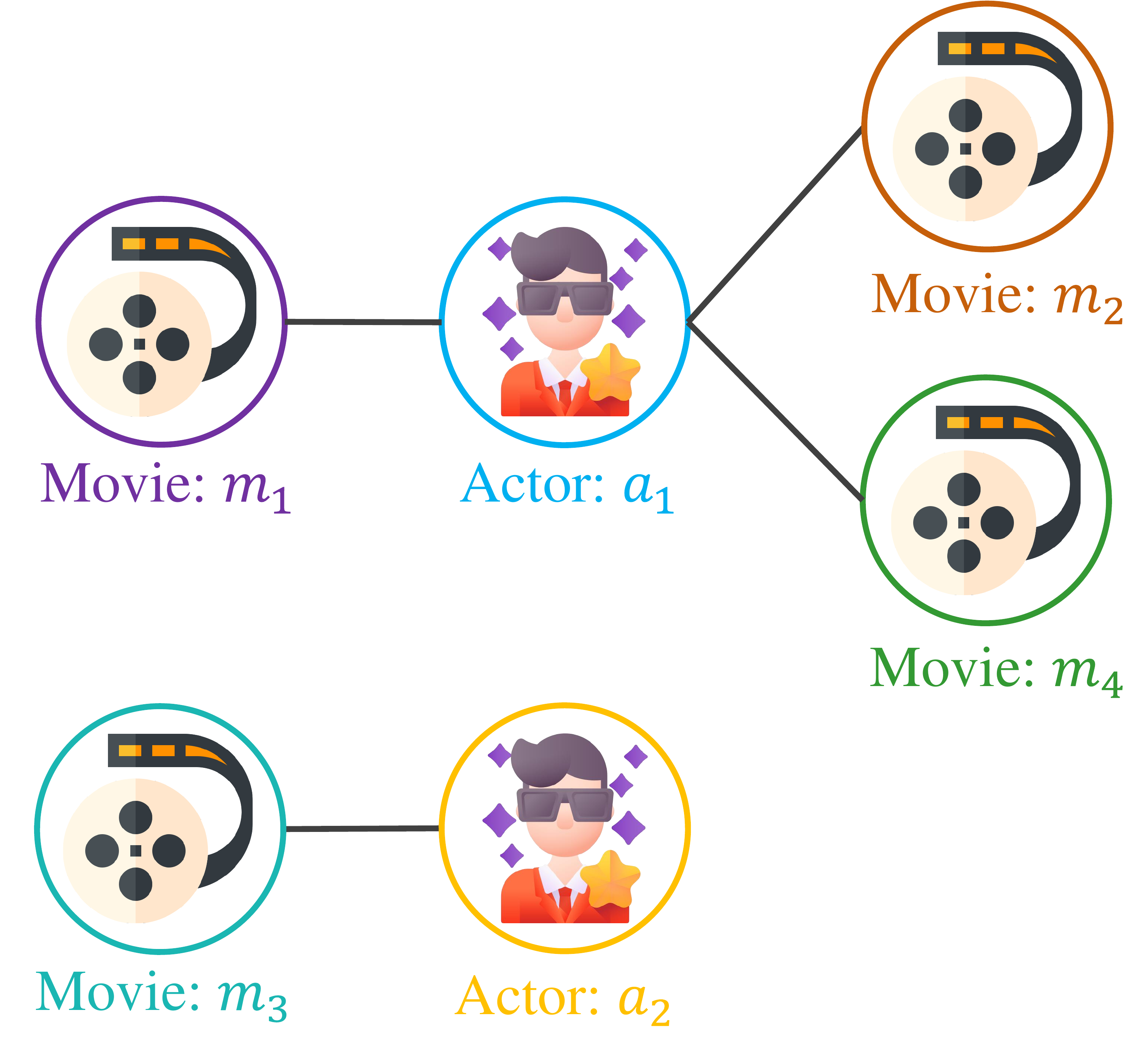}
		\caption{The movie nodes with different node centralities.}
		\label{fig:example_left}
	\end{subfigure}
	\hfill
	\hspace{.1in}
	\begin{subfigure}[b]{0.22\textwidth}
		\centering
		\includegraphics[width=\textwidth]{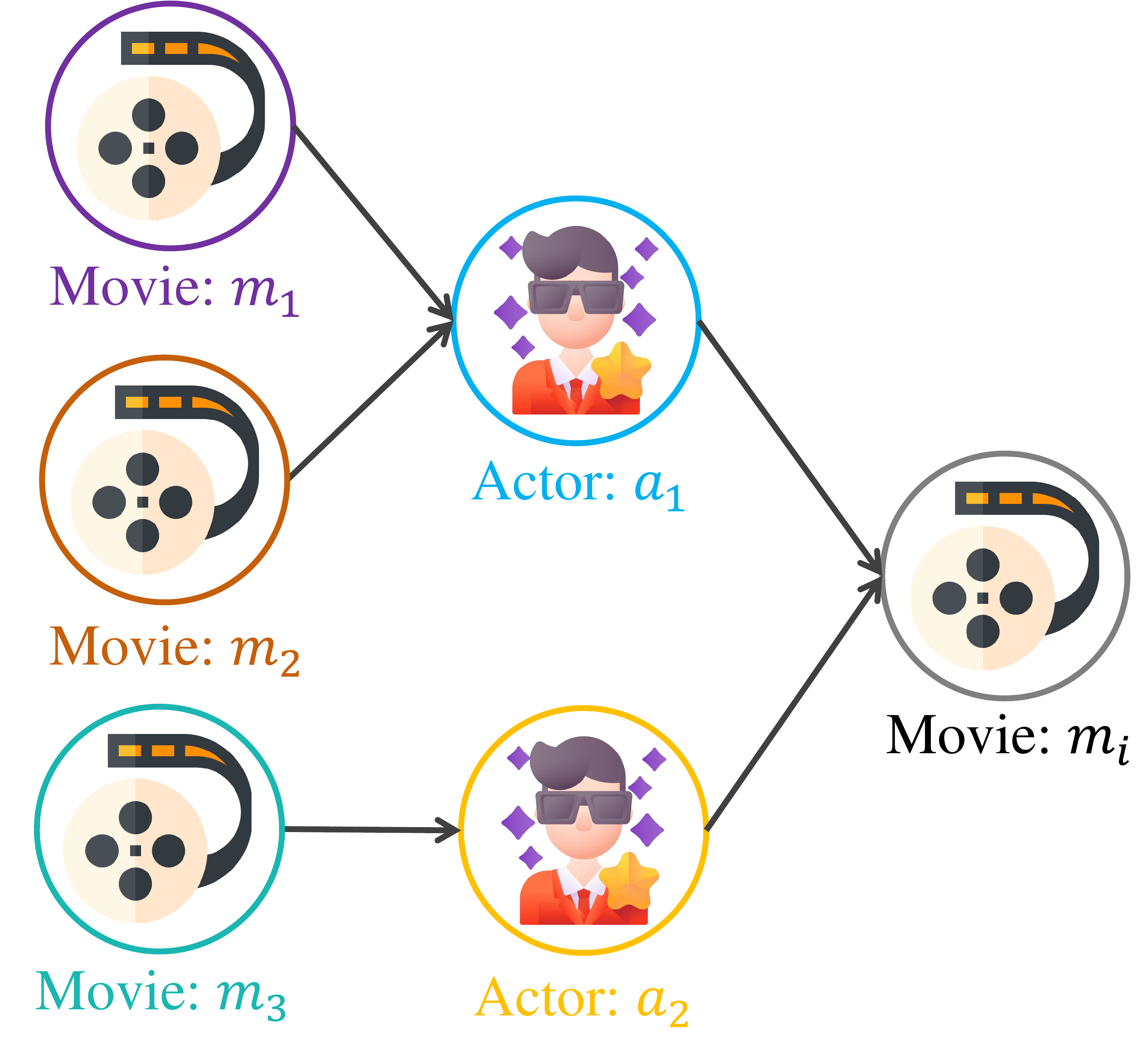}
		\caption{Information aggregation on different graph structures.}
		\label{fig:example_right}
	\end{subfigure}
	\vspace{-.1in}
	\caption{The heterogeneous graph in different structures.}
	\vspace{-.15in}
	\label{fig:aggregation_example}
\end{figure}

Most graph embedding methods assume that the input is a homogeneous graph, where all nodes are of the same type, and so are edges. For example, in drug-drug interaction prediction, all nodes are drugs, and all edges are binary labels $0/1$, indicating whether taking two drugs together can lead to a significant difference.
However, many real-world graphs are heterogeneous, where both nodes and edges have different types. As shown in Figure \ref{fig:aggregation_example}, if there is a network about actors and movies, there are two different types of nodes (i.e., actors, movies).
A heterogeneous graph usually contains richer information than the corresponding homogeneous graph, making it harder to encode. 
Previously, there are some unsupervised  methods~\cite{DBLP:journals/corr/ShangQLKHP16,Dong:2017:MSR:3097983.3098036,Fu:2017:HEM:3132847.3132953,shi2018heterogeneous} and  semi-supervised methods~\cite{Wang:2019:HGA:3308558.3313562,yun2019graph,hu2020heterogeneous,fu2020magnn} to learning the embeddings of heterogeneous graphs. A common part of the above work is that they are built upon meta-paths.
A meta-path is an ordered sequence of node types and edge types defined on the network schema, which describes a composite relation between the nodes types involved. 
For example, in the IMDB dataset~\cite{fu2020magnn}, which consists of a heterogeneous graph with three types of nodes: movies, actors, and directors, Movie-Actor-Movie (M-A-M) and Movie-Director-Movie (M-D-M) are two different meta-paths. 
The M-A-M meta-path describes the co-actor relation between two movies, and the M-D-M meta-path associates two movies directed by the same director. 
Depending on the meta-paths, the relation between nodes in the heterogeneous graph can have different semantics. 
Most of the existing heterogeneous graph neural networks utilize the meta-paths to sample meta-path based neighbor nodes and aggregate the node information to learn the heterogeneous graph's representations.

\xhdr{Motivation}
However, most previous heterogeneous graph embedding methods overlook the graph structure information when learning the embeddings.
Firstly, previous methods overlook the node centrality information determined by the graph structure, which measures how important a node is in the graph.
For instance, in Figure~\ref{fig:example_left}, both of the Movie nodes $m_1$ and $m_3$ connect with one Actor node only, which are  $a_1$ and $a_2$, respectively. 
The actor $a_1$ is more famous since he performs many other movies, so the movie $m_1$ performed by $a_1$ should be more popular than the movie $m_2$ performed by actor $a_2$. However, existing heterogeneous graph embedding methods cannot model this type of structural information that measures a node's importance.
Secondly, most of the existing meta-path based methods assume that each meta-path instance is isolate. They do not consider the graph structure when they aggregate information on the meta-path.
For instance, there are three meta-path instances $m_1-a_1-m_i$, $m_2-a_1-m_i$ and $m_3-a_2-m_i$ in Figure~\ref{fig:example_right}. Most of the previous methods~\cite{Wang:2019:HGA:3308558.3313562,fu2020magnn} assume these three instances are isolate, and combine them via some techniques like attention mechanism to aggregate the information of $m_1$, $m_2$ and $m_3$ to node $m_i$.
However, the meta-path instances $m_1-a_1-m_i$ and $m_2-a_1-m_i$ share the intermediate node $a_1$, and the nodes $m_1$ and $m_2$ should have a closer connection than the node $m_3$.
Therefore, we should consider these differences caused by the graph structure when aggregating information to node $m_i$.


\xhdr{Our method}
To incorporate the structure information into the graph embeddings, we propose a new model, Structure-Aware Heterogeneous Graph Neural Network (SHGNN), for heterogeneous graph embedding.
SHGNN first uses an aggregation module to provide initial embedding of all nodes to aggregate the features (Section~\ref{subsec:node_structural}).
To encode the node centrality, we utilize the coverage times of meta-path instances for representing the node centrality (Section~\ref{subsec:centrality}).
To incorporate the graph structure information into the information aggregation on meta-path, SHGNN applies a tree-attention aggregator to aggregate information for every meta-path (Section~\ref{subsec:meta-path}).
In this way, SHGNN can capture the overall graph structure information instead of processing each meta-path instance individually.
Finally, we utilize a meta-paths aggregator to fuse the information aggregated from different meta-paths. Thus we can learn the comprehensive semantics ingrained in the heterogeneous graph.
We conduct downstream node classification and node clustering tasks on several benchmark datasets to evaluate our proposed structure-aware heterogeneous graph neural network. The experimental results show that our SHGNN outperforms existing heterogeneous graph embedding methods.

\xhdr{Contributions}
Our contributions are summarized as follows:
\begin{enumerate}[leftmargin=1.5em]
	\item We explicitly model the importance of nodes using node centrality. Compared with previous work on homogeneous graphs, which simply uses the degrees of a node, we leverage the coverage times of meta-path instances to differentiate the importance of nodes in different graph structures.
	
	
	\item 
	We use a tree-based attention module to aggregate information on a specific meta-path and meanwhile consider the graph structure in multiple meta-path instances. Next, we use another attention module to aggregate the information across different types of meta-paths. In this way, the structural information in the meta-paths is fully leveraged.
	
	\item  We conduct experiments on three IMDB, DBLP and ACM datasets. Our method achieved state-of-the-art results on node classification and clustering tasks on all above datasets. We also provide extensive studies to show how our method works.
	
\end{enumerate}

\section{Related Work}

\begin{figure*}[t]
	\centering
	\begin{subfigure}[b]{0.24\textwidth}
		\centering
		\includegraphics[width=\textwidth]{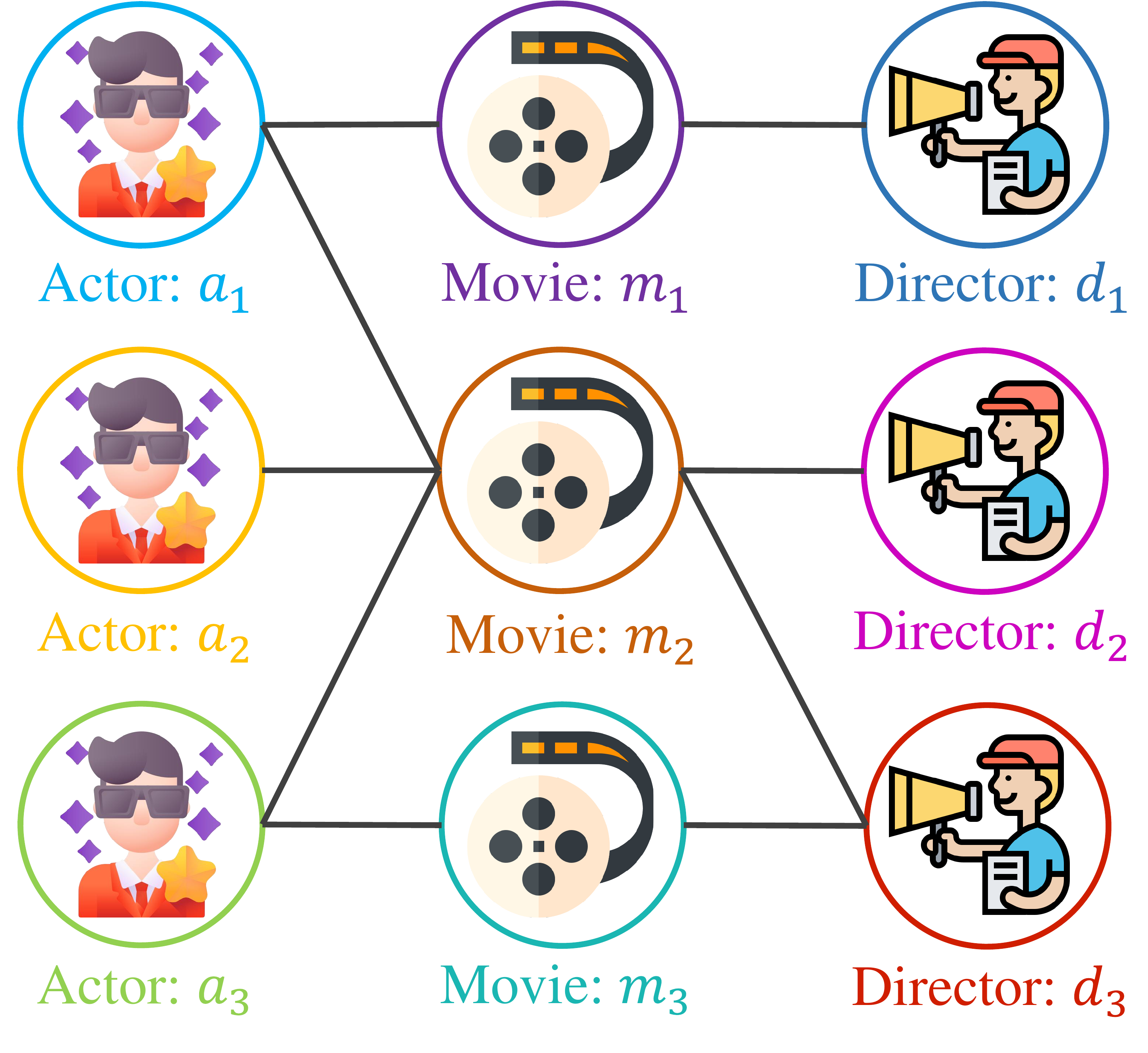}
		\caption{Heterogeneous Graph}
		\label{fig:het-graph}
	\end{subfigure} 
	\hfill
	\begin{subfigure}[b]{0.24\textwidth}
		\centering
		\includegraphics[width=\textwidth]{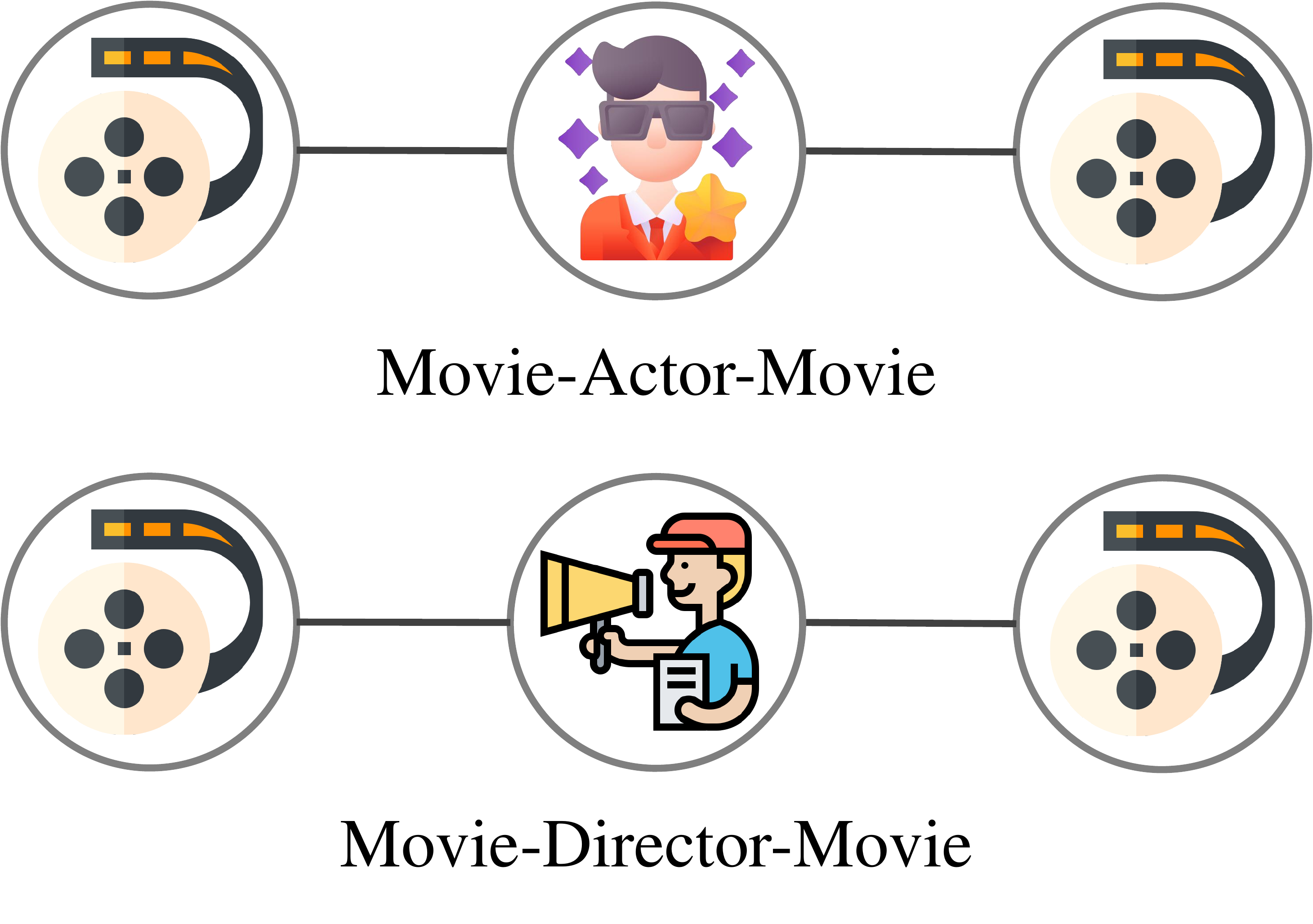}
		\caption{Meta-paths}
		\label{fig:metapath}
	\end{subfigure}
	\hfill
	\begin{subfigure}[b]{0.24\textwidth}
		\centering
		\includegraphics[width=\textwidth]{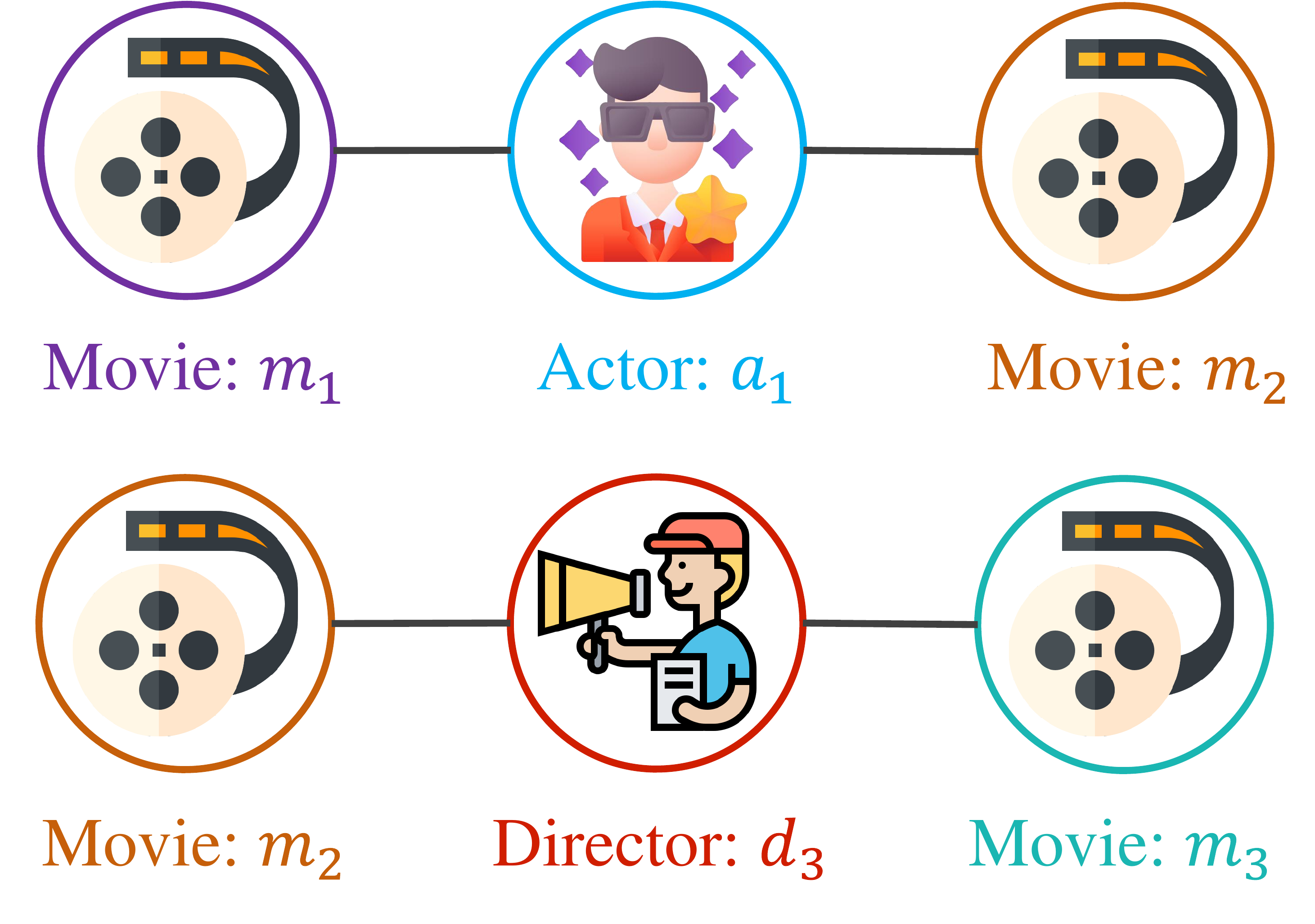}
		\caption{Meta-path Instances}
		\label{fig:metapath_instance}
	\end{subfigure}
	\hfill
	\begin{subfigure}[b]{0.24\textwidth}
		\centering
		\includegraphics[width=\textwidth]{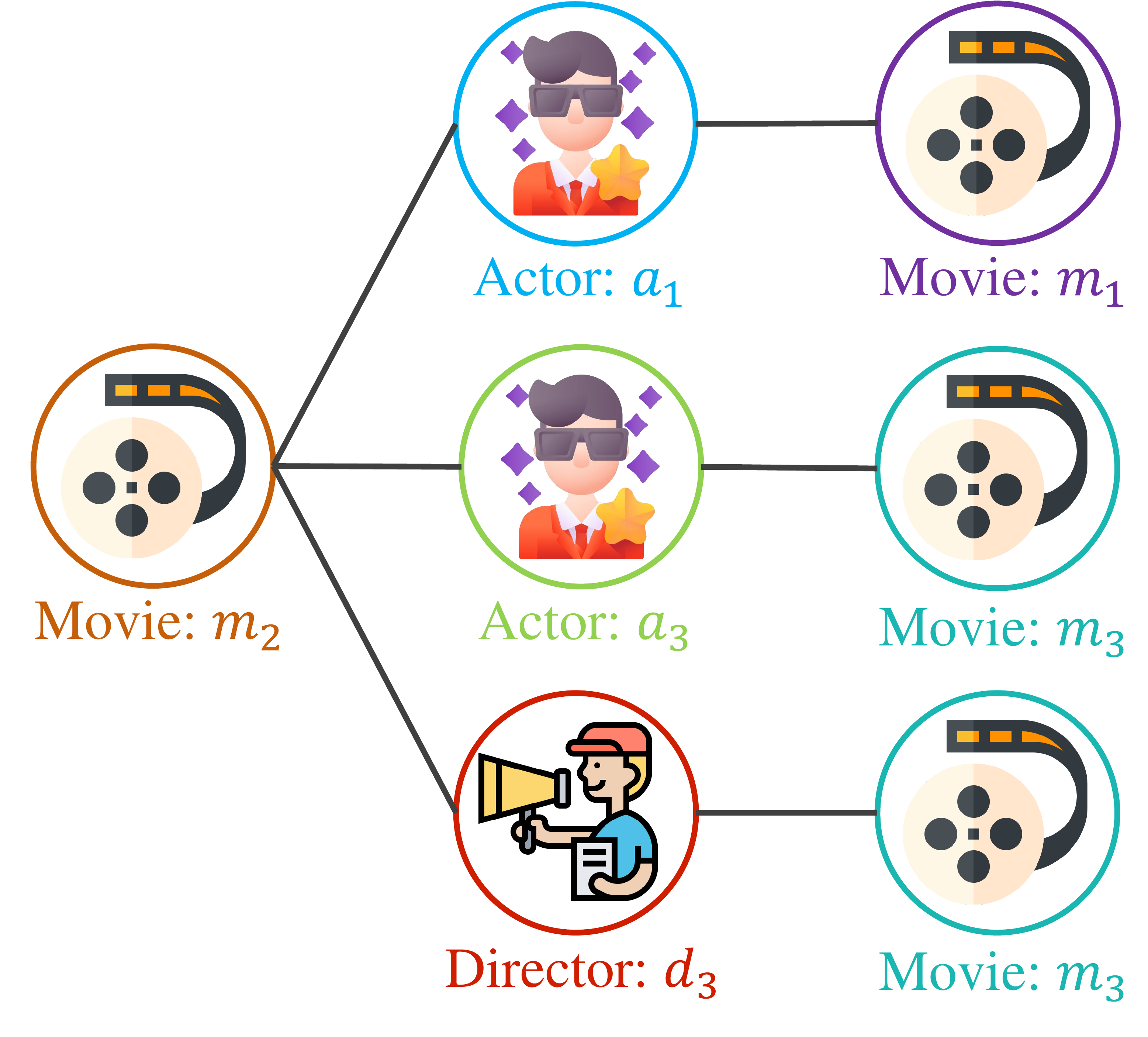}
		\caption{Meta-path based Neighbors}
		\label{fig:metapath-neighbor}
	\end{subfigure}
		\vspace{-.1in}
	\caption{The illustrations of some examples in Section~\ref{sec:preliminary}.}
		\vspace{-.15in}
	\label{fig:illustration}
\end{figure*}

\subsection{Graph Embedding}
\label{sec:GNN_survey}
Graph embedding aims to learn the low dimensional representations of graph nodes while preserving the graph structure and property so that the graph embeddings can apply to the downstream tasks. 
We can further divide graph embedding into two types: homogeneous and heterogeneous graph embedding.

Among the homogeneous graph embedding methods, the LINE~\cite{Tang:2015:LLI:2736277.2741093} exploits the first-order and second-order proximity between nodes to learn node embeddings.
The DeepWalk~\cite{Perozzi:2014:DOL:2623330.2623732}, node2vec~\cite{Grover:2016:NSF:2939672.2939754}, TADW~\cite{Yang:2015:NRL:2832415.2832542}, and Struc2vec~\cite{ribeiro2017struc2vec} are random walk based methods, which utilize random walk to generate node sequence and feed the sequence to a skip-gram~\cite{mikolov2013efficient} model learn node embeddings.
SDNE~\cite{wang2016structural} implies multiple layers of non-linear functions to capture highly non-linear network structure.

Heterogeneous graph embedding needs to further operate on nodes and edges that have different types.
Metapath2vec~\cite{Dong:2017:MSR:3097983.3098036} generates random walks guided by a single meta-path, and feeds it to a skip-gram model~\cite{DBLP:journals/corr/abs-1301-3781} to generate node embeddings.
ESim~\cite{DBLP:journals/corr/ShangQLKHP16} uses the user-defined meta-paths to learn the node embeddings from sampled positive and negative meta-path instances.
HIN2vec~\cite{Fu:2017:HEM:3132847.3132953} leverages multiple prediction training tasks to learn representations of nodes and meta-paths of a heterogeneous graph.
HERec~\cite{shi2018heterogeneous} translates a heterogeneous graph into a homogeneous graph based on meta-path based neighbors and uses the DeepWalk model to learn the node embeddings of the specific type.

\subsection{Graph Neural Network}
\label{sec:heter_survey}
The graph neural network (GNN) utilizes the deep neural network to learn the node embeddings via message passing.
Based on graph signal processing, spectral-based GNNs were first proposed to perform graph convolution in the Fourier domain of a graph.
ChebNet~\cite{Defferrard:2016:CNN:3157382.3157527} uses Chebyshev polynomials to filter graph signals (node features) in the graph Fourier domain. 
The graph convolutional network (GCN)~\cite{DBLP:conf/iclr/KipfW17} constrains and simplifies the parameters of ChebNet to address the overfitting problem and improve the performance.
On the other hand, spatial-based GNNs utilize the aggregator to aggregate information from neighbors for each node, imitating the convolution operations of convolutional neural networks for image data.
GraphSAGE~\cite{DBLP:conf/nips/HamiltonYL17} proposes the general format of aggregator functions for efficient learning of node embeddings.
Motivated by the Transformer~\cite{DBLP:conf/nips/VaswaniSPUJGKP17}, GAT~\cite{DBLP:conf/iclr/VelickovicCCRLB18} leverage the self-attention mechanism to calculate the relative importance of each neighbor's information in the aggregator.
GGNN~\cite{DBLP:journals/corr/LiTBZ15} adds a gated recurrent unit (GRU)~\cite{DBLP:journals/corr/ChoMGBSB14} to the aggregator by treating the aggregated neighborhood information as the input to the GRU of the current time step.
GaAN~\cite{DBLP:conf/uai/ZhangSXMKY18} combines GRU with the gated multi-head attention mechanism for spatiotemporal graphs.

The GNNs mentioned above operate on node features in the same embedding space, designing for homogeneous graphs. Therefore, they can not naturally apply to heterogeneous graphs with different nodes and edges.
To process the heterogeneous graph, some heterogeneous GNNs were proposed.
HAN~\cite{Wang:2019:HGA:3308558.3313562} converts a heterogeneous graph into multiple meta-path connected homogeneous graphs and uses a graph attention network architecture to aggregate information from the meta-path based neighbors and leverages the attention mechanism to combine various meta-paths.
MAGNN~\cite{fu2020magnn} utilizes the intra-meta-path aggregation to incorporate the word embedding of intermediate nodes and the inter-meta-path aggregation to combine messages from multiple meta-paths.
MAGNN-AC-\cite{jin2021heterogeneous} use topological relationship between nodes as guidance to complete attributes for no-attribute nodes.

However, the heterogeneous GNNs introduced above have the limitations of ignoring the node centrality information determined by the graph structure, or overlooking the graph structure when aggregating information on the meta-path.
Although they might have improved the performance of homogeneous graph embedding methods for some heterogeneous graph datasets, there is still room for improvement by incorporating the graph structure information into the heterogeneous graph embeddings.

\section{PRELIMINARY}
\label{sec:preliminary}
In this section, we will introduce some definitions and the corresponding examples in the heterogeneous graph.
\begin{Definition}{\textbf{Heterogeneous Graph~\cite{sun2013mining}.}}
	A heterogeneous graph is a graph $\mathcal{G}=\left(\mathcal{V},\mathcal{E}\right)$, where $\mathcal{V}$ is the set of nodes and $\mathcal{E}$ is the set of edges. It associates with a node type mapping function $\phi : \mathcal{V} \rightarrow \mathcal{A}$, and an edge type mapping function $\psi : \mathcal{E} \rightarrow \mathcal{R}$, where $|\mathcal{A}|+|\mathcal{R}|>2$. The $\mathcal{A}$ and $\mathcal{R}$ are the predefined sets of node types and edge types in the heterogeneous graph, respectively.
\end{Definition}
\begin{Example}
Figure~\ref{fig:het-graph} is a example of heterogeneous graph with three node types (Actors, Movies, and Directors) and two edge types (Actor-Movie edge and Movie-Director edge).
\end{Example}

\begin{Definition}{\textbf{Meta-path~\cite{sun2011pathsim}.}}
	A meta-path $P$ is a path in the form of $A_{1} \stackrel{R_{1}}{\longrightarrow} A_{2} \stackrel{R_{2}}{\longrightarrow} \cdots \stackrel{R_{l}}{\longrightarrow} A_{l+1}$ (abbreviated as $A_{1}-A_{2}-\cdots A_{l+1}$), representing a composite relation $R=R_{1} \circ R_{2} \circ \cdots \circ R_{l}$ between the node types $A_1$ and $A_{l+1}$. The symbol $\circ$ is the composition operator on relations.
\end{Definition}

\begin{Example}
Figure~\ref{fig:metapath} shows the Movie-Actor-Movie (M-A-M) and Movie-Director-Movie (M-D-M) meta-paths in the heterogeneous graph given in Figure~\ref{fig:het-graph}. Different meta-paths represent different semantics. For instance, meta-path M-A-M represents the co-actor relation, while M-D-M means they are directed by the same director.
\end{Example}

\begin{Definition}{\textbf{Meta-path Instance~\cite{fu2020magnn}.}}
	A meta-path instance $p$ of  a corresponding meta-path $P$ is a node sequence in the graph that follows the schema defined by $P$.
\end{Definition}

\begin{Example}
Figure~\ref{fig:metapath_instance} shows examples of some meta-path instances in Figure~\ref{fig:het-graph}. The meta-path instances $m_1-a_1-m_2$ and $m_2-d_3-m_3$ are corresponding to the meta-paths Movie-Actor-Movie and Movie-Director-Movie, respectively.
\end{Example}

\begin{Definition}{\textbf{Meta-path based Neighbor~\cite{Wang:2019:HGA:3308558.3313562}.}}
The meta-path based neighbors $\mathcal{N}^P_v$ of a node $v$ under the meta-path $P$ are the set of nodes that connect with node $v$ via the meta-path instances of $P$. A neighbor connected by two different meta-path instances is regarded as two different nodes in $\mathcal{N}^P_v$. When the meta-path $P$ is symmetric, the set $\mathcal{N}^P_v$ also includes node $v$. 
\end{Definition}

\begin{Example}
Figure~\ref{fig:metapath-neighbor} shows some meta-path based neighbors of Movie $m_2$. Given the meta-path Movie-Actor-Movie, the meta-path based neighbors of $m_2$ are the $m_1$ and $m_3$. Similarly, the neighbor of $m_2$ based on meta-path Movie-Director-Movie is $m_3$. We can get meta-path based neighbors by the multiplication of a sequences of adjacency matrices.
\end{Example}


\begin{Definition}{\textbf{Heterogeneous Graph Embedding~\cite{fu2020magnn}.}}
	Given a heterogeneous graph $\mathcal{G}=\left(\mathcal{V},\mathcal{E}\right)$, the heterogeneous graph embedding task is to learn the $d$-dimensional node representations $\mathbf{h}_{v} \in \mathbb{R}^{d}$ for all $v \in \mathcal{V}$ with $d \ll |\mathcal{V}|$ that are able to capture rich structural and semantic information involved in $\mathcal{G}$.
\end{Definition}
\section{Approach}
\label{sec:method}
\begin{figure*}[t]
	\centering
	\includegraphics[width=0.98\textwidth]{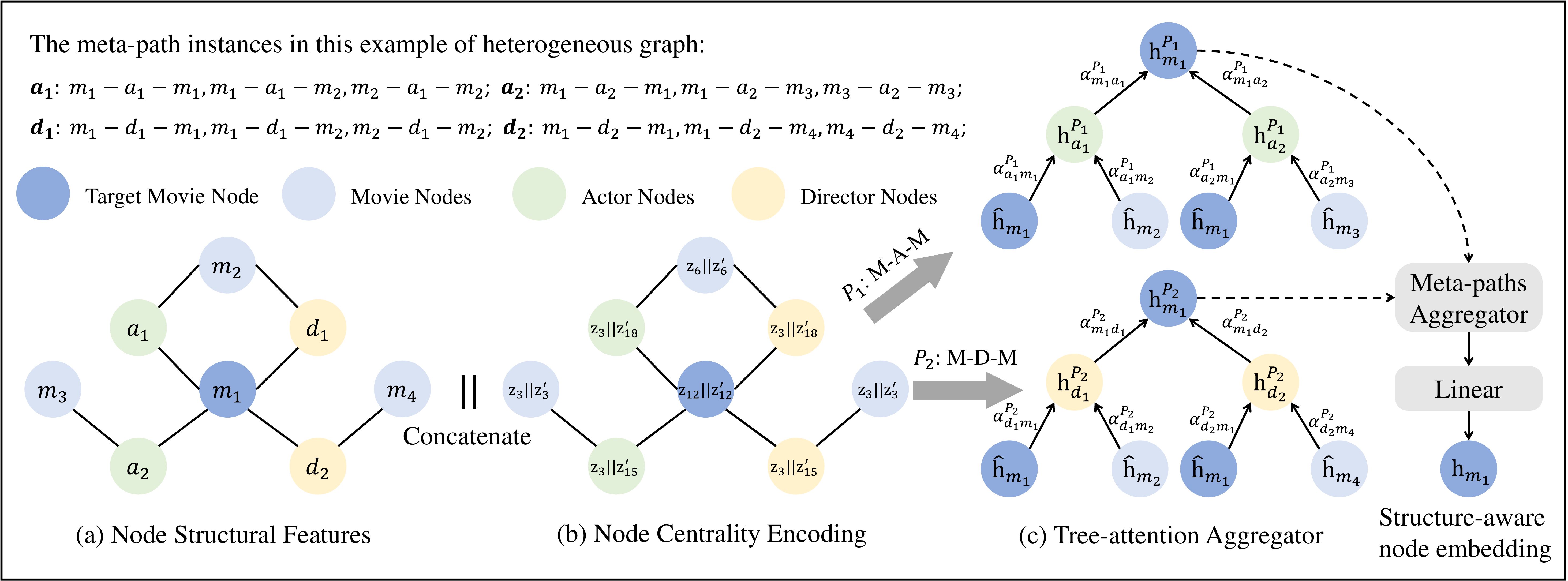}
	\vspace{-0.1in}
	\caption{The illustration of our structure-aware graph neural network on the heterogeneous graph of IMDB.}
	\vspace{-0.15in}
	\label{fig:framework}
\end{figure*}

In this section, we elaborate on our structure-aware heterogeneous graph neural network (SHGNN). 
Figure~\ref{fig:framework} is the illustration of our SHGNN on a heterogeneous graph example of IMDB. 
The key insights of our SHGNN are to consider the graph structural information when learning the embeddings, including the node structural features in Section~\ref{subsec:node_structural}, the node centrality encoding in Section~\ref{subsec:centrality} and the meta-path structural information encoder in Section~\ref{subsec:meta-path}.

\subsection{Node Structural Features}
\label{subsec:node_structural}
\begin{figure}[t]
	\centering
	\begin{subfigure}[b]{0.23\textwidth}
		\centering
		\includegraphics[width=\textwidth]{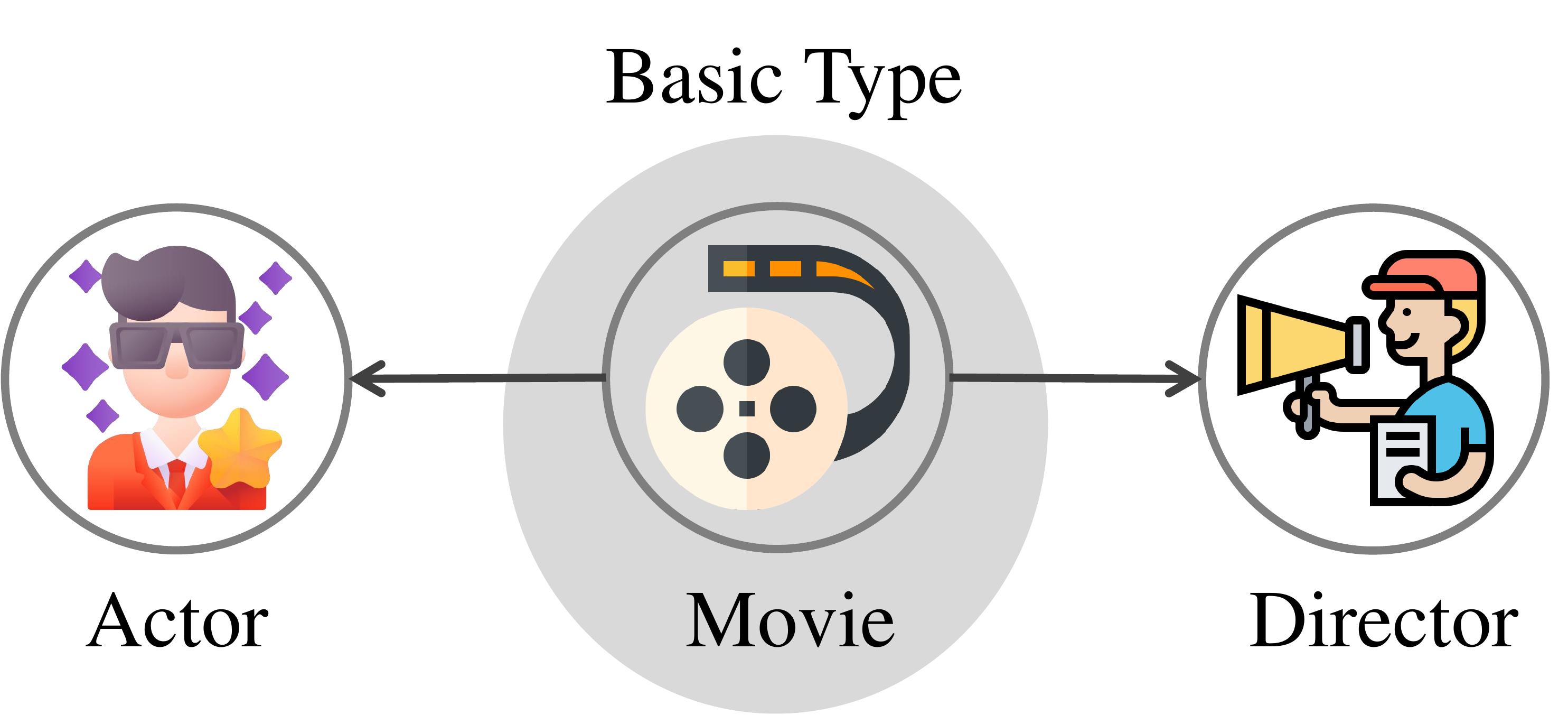}
		\caption{IMDB}
	\end{subfigure}
	\hfill
	\begin{subfigure}[b]{0.23\textwidth}
		\centering
		\includegraphics[width=\textwidth]{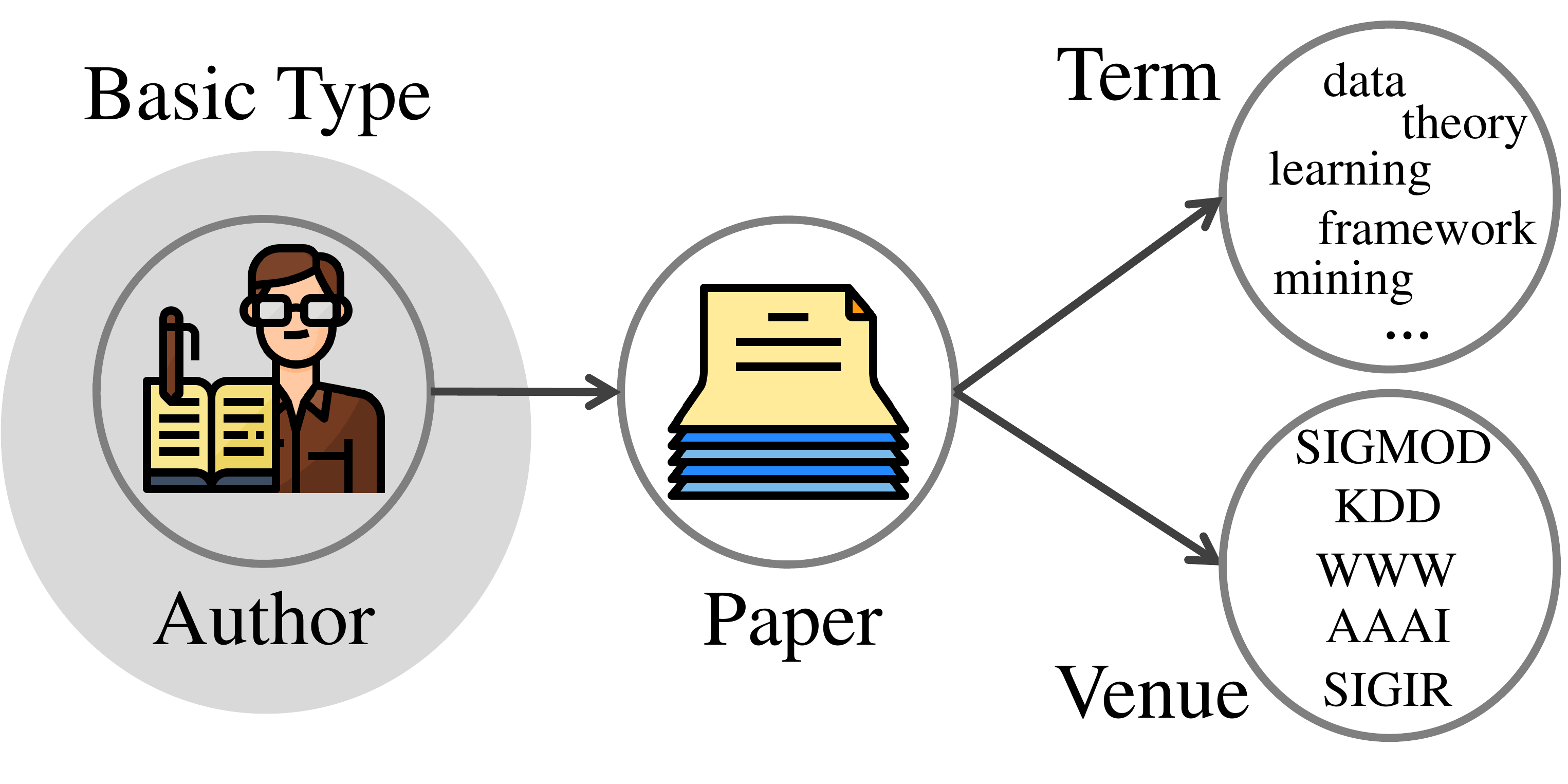}
		\caption{DBLP}
	\end{subfigure}
	\vspace{-.1in}
	\caption{Feature propagation on the network schema.}
	\vspace{-.15in}
	\label{fig:feature_propagation}
\end{figure}

\xhdr{Feature Propagation}
As pointed out in~\cite{jin2021heterogeneous}, many types of nodes in heterogeneous graphs are not adequately modeled, where their features are the corresponding word embeddings of the descriptions (i.e., whom the actor is, a brief description of a director, etc.). Such features cannot correctly model the graph due to the lack of structural information.
To represent the graph semantic information in the node features, we use a feature propagation module to learn the semantic information of nodes.
We first choose a type of nodes as the \textbf{basic type} (e.g., the Movie nodes in IMDB, the Author nodes in DBLP, and the Paper nodes in ACM), and initialize the nodes in the basic type with bag-of-words embeddings. The embeddings of the basic type's nodes are also the nodes' features.
Then as shown in Figure~\ref{fig:feature_propagation}, we propagate the features from the basic type's nodes to the nodes in other types on the network schema of a heterogeneous graph. For a node $v \in \mathcal{V}_{A_i}$ of type $A_i \in \mathcal{A}$, and the nodes in type $A_i$ do not have features, we obtain the node $v$'s features by aggregating the features from the nodes in type $A_j$:
\begin{equation}
	\mathbf{x}_{v}^{A_i} =\frac{1}{| \mathcal{N}_v^{ij}|} \sum_{k \in \mathcal{N}_v^{ij}}  \mathbf{x}_{k}^{A_j},
\end{equation}
where the nodes in the type $A_j$ has features and the type $A_j$ is the nearest type to type $A_i$ on the network schema. 
$\mathcal{N}_v^{ij}$ are the neighbors of node $v$ that in type $A_j$, and $\mathbf{x}_{k}^{A_j}$ is the feature of the node $k$, which is one of the node $v$'s neighbors in type $A_j$.
Taking the node $a_1$ (type Actor) in Figure~\ref{fig:het-graph} as an example, the type Movie is the basic type, and the node $a_1$ would aggregate its feature from the node $m_1$ and $m_2$. After the feature propagation, each node in the heterogeneous graph has the local structural embedding, which involves semantic information. The embeddings of an Actor node can reflect all the participated movies.
Therefore, the feature propagation module can incorporate a node's graph structural information into the node feature.

\xhdr{Type-specific Transformation}
Due to the heterogeneity of nodes, different types of nodes have different feature spaces. Hence, we apply a type-specific linear transformation for each type of nodes by projecting their features into the same latent factor space.  
For a node $v \in \mathcal{V}_{A_i}$ of type $A_i \in \mathcal{A}$, we have:
\begin{equation}
	\mathbf{\hat{x}}_{v}^{A_i}=W^x_{A_i}\mathbf{x}_{v}^{A_i},
\end{equation}
where $\mathbf{\hat{x}}_{v}^{A_i} \in \mathbb{R}^{d_1}$  is the projected latent vector of node $v$. $W_{A_i} \in \mathbb{R}^{d_1 \times d_0}$ is the parametric weight matrix for type $A_i$'s nodes.
The type-specific transformation addresses the heterogeneity of a graph.

\subsection{Node Centrality Encoding}
\label{subsec:centrality}
The node centrality, measuring how important a node is in the graph, is usually significant structural information for graph understanding. For instance, the celebrities who have massive followers are vital factors in predicting the social network trend~\cite{marshall2010promotion}.
In the homogeneous graph, some existing methods~\cite{ying2021transformers} would utilize the degree centrality, including the indegree and outdegree of a node to measure a node's centrality.
However, we found that the degree centrality is not distinguishable for the heterogeneous graph's nodes since many nodes have the same degree. 
For example, in the heterogeneous graph of IMDB dataset~\cite{fu2020magnn}, 99.91\% of the Movie nodes have the same degree 4.

To properly model the node centrality in the heterogeneous graph, we propose utilizing the coverage times of meta-path instances to represent the node centrality. 
Specifically, we count the number of times $c(v)$ that a node $v$ covered by all the meta-path instances in the heterogeneous graph. 
When more meta-path instances cover a node, we assume that this node is more important in the graph and has a higher $c(v)$ value.
For example, in the heterogeneous graph of IMDB, even two Movie nodes have the same number of degrees (actors), the Movie node covered by more meta-path instances (Movie-Actor-Movie) indicates that this movie is performed by the actors performing more other movies, so this movie is more important than the other movie covered by fewer meta-path instances.
After that, we assign each node a one hot embedding vector $\mathbf{z}_{c(v)}$ according to the ${\tiny }c(v)$ value.

Moreover, to capture more node centrality information, we further count the number of times $c^+(v)$ that a node $v$ as the direct neighbor of any one of the nodes in a meta-path instance. 
We can simply calculate the $c^+(v)$ with the following equation:
\begin{equation}
	c^+(v)=\sum_{k\in \mathcal{N}_v} c(k),
\end{equation}
where $\mathcal{N}_v$ is the set of all direct neighbors of node $v$, and $c(k)$ is the number of times that meta-path instances cover node $k$.
We assign another one hot embedding vector $\mathbf{z}'_{c^+(v)}$ according to the $c^+(v)$ value. 
As the centrality encodings $\mathbf{z}_{c(v)}$ and $\mathbf{z}'_{c^+(v)}$ are applied to each node, we concatenate them and apply another type-specific transformation on them:
\begin{equation}
	\mathbf{\hat{z}}_v^{A_i} = W^z_{A_i}\left( \mathbf{z}_{c(v)}|| \mathbf{z}'_{c^+(v)}\right),
\end{equation}
where $\mathbf{z}_{c(v)}\in \mathbb{R}^{d_1}$ and $\mathbf{z}'_{c^+(v)}\in \mathbb{R}^{d_1}$ are one hot embedding vectors specific by the counts $c(v)$ and $c^+(v)$, respectively. The $W^z_{A_i}\in \mathbb{R}^{d_1 \times 2d_1}$ is the weighted transform matrix for type $A_i$, and $\mathbf{\hat{z}}_v^{A_i}$ is the transformed centrality embedding of node $v$.

We concatenate the node structural features $\mathbf{\hat{x}}_{v}^{A_i}$ and node centrality embedding $\mathbf{\hat{z}}_v^{A_i}$ as the node $v$'s latent vector $\mathbf{\hat{h}}_v$ ($\mathbf{\hat{h}}_v \in \mathbb{R}^{2d_1}$).
\begin{equation}
	\mathbf{\hat{h}}_v = \left(\mathbf{\hat{x}}_{v}^{A_i} || \mathbf{\hat{z}}_v^{A_i}\right).
\end{equation}

\subsection{Meta-path Structural Information Encoder}
\label{subsec:meta-path}
After obtaining the latent vector of each node, we utilize a meta-path structural information encoder to aggregate the latent vectors of meta-path based neighbors. The proposed encoder includes a tree-attention aggregator for incorporating the graph structure into the information aggregation on the meta-path, and a meta-path aggregator to fuse the information aggregated by various meta-paths in the heterogeneous graph.

\xhdr{Tree-attention Aggregator}
To incorporate the graph structure into the information aggregation on the meta-path, we propose a tree-attention aggregator to aggregate the information for a meta-path. 
Given a meta-path $P_j$ and a target node $t$ whose type is the end of $P_j$, we expand all the $P_j$'s meta-path instances that end with the node $t$ as a rooted tree. This tree's root node is the target node $t$, and its leaf nodes are the meta-path based neighbors of the target node $t$.
Then we perform a tree-attention to aggregate the information from the tree's leaf nodes to the root node.
For a parent node $v$, since different child nodes would differently contribute to the parent node, we compute the cosine similarity between the node $v$'s latent vector $\mathbf{\hat{h}}_v$ and its child nodes' latent vectors as the attention values from the child nodes to the parent node:
\begin{equation}
	e^{P_j}_{vu} = \mathrm{Cosine}(\mathbf{\hat{h}}_v, 	\mathbf{\hat{h}}_k)  =   \dfrac{\mathbf{\hat{h}}_v \cdot \mathbf{\hat{h}}_k}{|| \mathbf{\hat{h}}_v || \cdot || \mathbf{\hat{h}}_k ||}, k \in \mathcal{C}_{v}^{P_j},
\end{equation}
where $\mathcal{C}_{v}^{P_j}$ is the child node set of the node $v$ on the tree generated by meta-path the $P_j$. 
Hence, when the child node's latent vector $\mathbf{\hat{h}}_k$ has a higher similarity with parent node's vector $\mathbf{\hat{h}}_v$, the child node $k$ would have a higher attention weight to the parent node $v$.
We normalize the attention value $e^{P_j}_{vu}$ across all child nodes of node $v$ using the softmax function, then we can obtain the aggregated weight $\alpha_{v k}^{P_j}$ from the child node $k$ to the parent node $v$. We use the aggregated weights to aggregate the child nodes' information $\mathbf{h}_{k}^{P_j}$ to the parent node:
\begin{equation}
	\begin{aligned}
		\alpha_{v k}^{P_j} &= \frac{\exp \left(e_{v k}^{P_j}\right)}{\sum_{s \in \mathcal{C}_{v}^{P_j}} \exp \left(e_{v s}^{P_j}\right)},\\
		\mathbf{h}_{v}^{P_j} &= \sum_{k \in \mathcal{C}_{v}^{P_j}} \alpha_{v k}^{P_j} \cdot \mathbf{h}_{k}^{P_j}.
	\end{aligned}
	\vspace{-0.1in}
\end{equation}
Note that each parent node $v$ aggregate the information $\mathbf{h}_{v}^{P_j}$ from its child nodes, and if node $k$ is a leaf node in the tree, then $\mathbf{h}_{k}^{P_j}$ is equal to the node $k$'s latent vector $\mathbf{\hat{h}}_k$.
As shown in Figure~\ref{fig:framework}c, from leaf nodes to root node, the tree-attention aggregator step-by-step aggregates each parent node's information from its child nodes. Since the different child nodes and different graph structures would result in different aggregated results on the parent node, we can incorporate the graph structure into the information aggregation on the meta-path. 
Moreover, we do not introduce other parameters in the tree-attention aggregator, so we can model the effect of different graph structures without increasing parameters.

\xhdr{Meta-paths Aggregator}
After aggregating the node information on each meta-path's corresponding tree, we combine the aggregated information of each meta-path using an meta-paths aggregator. For a node type $A_i$, we have $|\mathcal{V}_{A_i}|$ sets of latent vectors: $\left\{\mathbf{h}_{v}^{P_{1}}, \mathbf{h}_{v}^{P_{2}}, \ldots, \mathbf{h}_{v}^{P_{M}}\right\}$ for $v\in \mathcal{V}_{A_i}$, where $M$ is the number of meta-paths for type $A_i$. 
Since different meta-paths are not equally important in a heterogeneous graph, we leverage the attention mechanism to assign different weights to different meta-paths. 
We firstly summarize each meta-path $P_{j}\in \mathcal{P}_{A_i}$ by averaging the transformed meta-path specific node vectors for all nodes $v \in \mathcal{V}_{A_i}$:
\begin{equation}
	\mathbf{s}_{P_{j}}=\frac{1}{|\mathcal{V}_{A_i}|} \sum_{v \in \mathcal{V}_{A_i}} \tanh \left(W^h_{A_i} \mathbf{h}_{v}^{P_{j}}+\mathbf{b}_{A_i}\right),
\end{equation}
where $W^h_{A_i}  \in \mathbb{R}^{d_1 \times 2d_1}$ and $\mathbf{b}_{A_i} \in \mathbb{R}^{d_1}$ are learnable parameters.
Then we use the attention mechanism to fuse the meta-path specific vectors of node $v$ as follows:
\begin{equation}
	\begin{aligned}
		u_{P_{j}}&=\mathbf{q}_{A_i}^{\intercal} \mathbf{s}_{P_{j}},\\
		\beta_{P_{j}}&=\frac{\exp \left(u_{P_{j}}\right)}{\sum_{P_k \in \mathcal{P}_{A_i}} \exp \left(u_{P_k}\right)}, \\
		\mathbf{h}_{v}^{\mathcal{P}_{A_i}} &= \sum_{P_j \in \mathcal{P}_{A_i}} \beta_{P_j} \cdot \mathbf{h}_{v}^{P_j},
	\end{aligned}
\end{equation}
where $\mathbf{q}_{A_i} \in \mathbb{R}^{d_1}$ is the parameterized vector for node type $A_i$. $\beta_{P_{j}}$ is the importance weight of meta-path $P_j$ to type $A_i$'s nodes, and $\mathbf{h}_{v}^{\mathcal{P}_{A_i}}$ is aggregated information of all meta-paths on node $v$.

\begin{algorithm}[t]
	\small
	\caption{The overall training process of our SHGNN.}
	\label{algo:SHGNN}
	\SetAlgoLined
	\KwIn{The heterogeneous graph $\mathcal{G}=\left(\mathcal{V},\mathcal{E}\right)$, 
		node types set $\mathcal{A}$, meta-paths set $\mathcal{P}$,
		features of basic type's node, and the number of layers $L$.
	}
	\KwOut{The node embeddings $\left\{\mathbf{h}_{v}, \forall v \in \mathcal{V}\right\}$.}
	
	\For{node type $A_i \in \mathcal{A}$}{
		\For{node $v \in \mathcal{V}_{A_i}$}{
			Feature Propagation $ \mathbf{x}_{v}^{A_i} \leftarrow \frac{1}{| \mathcal{N}_v^{ij}|} \sum_{k \in \mathcal{N}_v^{ij}}  \mathbf{x}_{k}^{A_j}$\;
			Type-specific transformation $\mathbf{\hat{x}}_{v} \leftarrow W^x_{A_i} \cdot \mathbf{x}_{v}^{A_i}$\;
			Count the times $c(v)$ and $c^+(v)$ for node $v$\;
			Node centrality encoding: $\mathbf{\hat{z}}_v \leftarrow W^z_{A_i}( \mathbf{z}_{c(v)}|| \mathbf{z}'_{c^+(v)})$\;
			Obtain the node latent vector: $\mathbf{\hat{h}}_v \leftarrow \left(\mathbf{\hat{x}}_{v} || \mathbf{\hat{z}}_v\right)$\;
		}
	}
	\For{$l = 1 \ldots L$}{
		\For{node type $A_i \in \mathcal{A}$}{
			\For{meta-path $P_j \in \mathcal{P}_{A_i}$}{
				\For{$v \in \mathcal{V}_{A_i}$}{
					Generate a rooted tree according to the target node $v$ and $P_j$'s meta-path instances\;
					Perform tree-attention $\left[\mathbf{h}_{v}^{P_j}\right]^{l} \leftarrow  \sigma\left(\sum_{k \in \mathcal{C}_{v}^{P_j}} \alpha_{v k}^{P_j} \cdot \left[\mathbf{h}_{k}^{P_j}\right]^{l}\right)$\;
				}
			}
			Calculate the weight $\beta_{P}$ for meta-path $P_j \in \mathcal{P}_{A_i}$\;
			Fuse the embeddings from different meta-paths $\left[\mathbf{h}_{v}^{\mathcal{P}_{A_i}}\right]^{l} \leftarrow \sum_{P_j \in \mathcal{P}_{A_i}} \beta_{P_j} \cdot \left[\mathbf{h}_{v}^{P_j}\right]^{l}, \forall v \in \mathcal{V}_{A_i}$\;
		}
		Layer output projection: $\mathbf{h}_{v}^{l} = \sigma\left(W_{o}^{l} \cdot \left[\mathbf{h}_{v}^{\mathcal{P}_{A_i}}\right]^{l}\right), \forall v \in \mathcal{V}_{A_i}$\;
	}	
\end{algorithm}

\subsection{Optimization Objective}
Finally, we use a linear layer with a ELU activation function to project the node embeddings $\mathbf{h}_{v}^{\mathcal{P}_{A_i}}$ to the desired output vector:
\begin{equation}
	\label{eq:output_proj}
	\mathbf{h}_{v} =\mathrm{ELU}\left(W_o \mathbf{h}_{v}^{\mathcal{P}_{A_i}}\right),
\end{equation}
where $W_o \in \mathbb{R}^{d \times 2d_1}$ is a weight matrix, and $\mathbf{h}_{v}$ is the structure-aware node representation in heterogeneous graph. The projection is task-specific, which can be a linear classifier for the node classification task or be a projection to the node similarity measuring space for the node clustering task.

We use the stochastic gradient descent algorithm (SGD) to train our framework and leverage the Adam~\cite{kingma2014adam} for tuning the learning rate.
We optimize our SHGNN by minimizing the cross entropy loss function $\mathcal{L}$:
\begin{equation} 
	\label{eq:semi-supervised-loss}
	\mathcal{L} = - \sum_{v \in \mathcal{V}_{L}} \sum_{c=1}^{C} \mathbf{y}_{v}[c] \cdot \log \mathbf{h}_{v}[c],
\end{equation}
where $\mathcal{V}_{L}$ is the set of nodes with labels, $C$ is the number of classes, $\mathbf{y}_{v}$ is the one-hot label vector of node $v$, and $\mathbf{h}_{v}$ is the predicted probability vector of node $v$. The Algorithm~\ref{algo:SHGNN} shows the overall training process of our SHGNN.

\section{Experiments}
\label{sec:experiment}
In this section, we study the efficacy of the proposed structure-aware heterogeneous graph neural network with experiments. We aim to answer the following research questions via the experiments:
\begin{itemize}[leftmargin=1.5em]
	\item \textbf{RQ1}: How does SHGNN perform in node classification task? 
	\item \textbf{RQ2}: How does SHGNN perform in node clustering task? 
	\item \textbf{RQ3}: What are the effect of different components in SHGNN ?
	\item \textbf{RQ4}: How do we understand the representation capability of different graph embedding methods? 
\end{itemize}

\subsection{Datasets}
We evaluate the performance of our proposed SHGNN framework on three benchmark heterogeneous graph datasets: IMDB, DBLP, and ACM. Table~\ref{tab:dataset} summarizes the statistics of the datasets. The detailed datasets' introductions are as follows:

\begin{itemize}[leftmargin=1.5em]
	\item \textbf{IMDB}\footnote{\url{https://www.imdb.com/}.} is an online database of movies and television programs. We use a subset of IMDB released by~\cite{fu2020magnn}, containing 4278 movies (M), 2081 directors (D), and 5257 actors (A) after extraction. Movies have one of three classes of labels (\textit{Action}, \textit{Comedy}, and \textit{Drama}) based on their genre information. The feature of each movie node is its plot keywords' bag-of-words representation. Following~\cite{fu2020magnn,jin2021heterogeneous}, we divide the movie nodes into training, validation, and testing sets of 400 (9.35\%), 400 (9.35\%), and 3478 (81.30\%) nodes, respectively.
	
	\item \textbf{DBLP}\footnote{\url{https://dblp.uni-trier.de/}.} is a computer science bibliography website. We use a subset of DBLP that also used in~\cite{fu2020magnn,jin2021heterogeneous}, containing 4057 authors (A), 14328 papers (P), 7723 terms (T), and 20 publication venues (V). The authors are divided into four research areas (\textit{Database}, \textit{Data Mining}, \textit{Artificial Intelligence}, and \textit{Information Retrieval}). The feature of each author node is the bag-of-words representation of their paper keywords. Following~\cite{fu2020magnn,jin2021heterogeneous}, we divide the author nodes into training, validation, and testing sets of 400 (9.86\%), 400 (9.86\%), and 3257 (80.28\%) nodes, respectively.
	\item \textbf{ACM}\footnote{\url{http://dl.acm.org/}.} is a collection of all ACM publications, including journals, conference proceedings, and books. 
	We use a subset of ACM extracted by~\cite{jin2021heterogeneous}, which comprises 4019 papers (P), 7167 authors (A), and 60 subjects (S).
	The papers are divided into three classes (\textit{Database}, \textit{Wireless Communication}, \textit{Data Mining}). 
	In this dataset, papers’ features are the bag-of-words representation of their keywords. 
	Following~\cite{jin2021heterogeneous}, we divide the paper nodes into training, validation, and testing sets of 400 (9.95\%), 400 (9.95\%), and 3219 (80.10\%) nodes, respectively.
\end{itemize}

\begin{table}[t]
	\centering
	\caption{Statistics of datasets.}
	\label{tab:dataset}
	\vspace{-0.15in}
	\resizebox{0.47\textwidth}{!}{
		\begin{tabular}{c|c|c|c}
			\hline
			\\[-1em]
			Datasets & Nodes & Edges & Meta-paths\\
			\\[-1em]
			\hline
			\\[-1em]
			IMDB & \specialcell{\# Movie (M): 4278 \\ \# Director (D): 2081 \\ \# Actor (A): 5257} & \specialcell{ \# M-D: 4278 \\ \#  M-A: 12828} & \specialcell{M-D-M \\ M-A-M}\\
			\\[-1em]
			\hline
			\\[-1em]
			DBLP & \specialcell{ \# Author (A): 4057 \\ \# Paper (P): 14328 \\ \# Term (T): 7723 \\  Venue (V): 20} & \specialcell{ \# A-P: 19645 \\  \# P-T: 85810 \\  \# P-V: 14328} & \specialcell{A-P-A \\ A-P-T-P-A \\ A-P-V-P-A}\\
			\\[-1em]
			\hline
			\\[-1em]
			ACM & \specialcell{\# Paper (P): 4019 \\ \# Author (A): 7167 \\ \# Subject (S): 60} & \specialcell{\# P-A: 13407 \\ \# P-S: 4019 } & \specialcell{P-A-P \\ P-S-P }\\
			[-1em]\\		
			\hline
	\end{tabular}}
	\vspace{-0.15in}
\end{table}

\begin{table*}[t]
	\caption{The node classification results (\%) on the IMDB, DBLP, and ACM datasets. }
	\vspace{-0.1in}
	\label{tab:node_class}
		\resizebox{0.8\textwidth}{!}{
	\begin{tabular}{c|c|c|c|c|c|c|c|c|c|c}
		\hline
		\\[-1em]
		\multirow{2}{*}{Dataset} & \multirow{2}{*}{Metrics} & \multirow{2}{*}{Train \%} &metapath  & \multirow{2}{*}{GCN}   & \multirow{2}{*}{GAT}  &\multirow{2}{*}{HetGNN} & \multirow{2}{*}{HAN}   & \multirow{2}{*}{MAGNN}  &MAGNN&   \multirow{2}{*}{\textbf{SHGNN}} \\
		&&  &  2vec&  & &&  &   &-AC& \\
		\\[-1em]
		\hline
		\\[-1em]
		\multirow{8}{*}{IMDB} & \multirow{4}{*}{Macro-F1} & 20\%       & 46.42        & 44.75 & 54.81 &48.92 & 57.61 & 58.11 &59.67& \textbf{60.69} \\ 
		&                           & 40\%       & 47.79         & 45.26 & 55.09 & 51.61 & 57.75 & 59.39 &60.18& \textbf{62.36}\\ 
		&                           & 60\%       & 48.25         & 46.72 & 55.71 & 53.00& 57.66 & 59.97 &60.60&  \textbf{63.23}\\
		&                           & 80\%       & 48.73         & 47.13 & 55.40 & 53.24& 57.23 & 60.02 &60.75&  \textbf{63.67}\\
		\\[-1em]
		\cline{2-11}
		\\[-1em]
		& \multirow{4}{*}{Micro-F1} & 20\%       & 48.08       & 47.44 & 55.02 &49.70 & 57.82 & 58.16 &59.84& \textbf{60.80}\\ 
		&                           & 40\%        & 49.55         & 47.62 & 55.56 & 55.29& 57.98 & 59.46 &60.38&  \textbf{62.48}\\ 
		&                           & 60\%       & 50.06         & 48.49 & 55.91 & 53.91& 57.87 & 60.05 &60.79&  \textbf{63.34}\\ 
		&                           & 80\%        & 50.68        & 48.73& 55.67 & 54.25 & 57.46 & 60.15 &60.98& \textbf{63.85} \\  	
		\\[-1em]
		\hline
		\\[-1em]
		\multirow{8}{*}{DBLP} & \multirow{4}{*}{Macro-F1} & 20\%      & 91.50    & 90.06 & 66.92&  91.72 & 91.69 & 92.53 &94.20& \textbf{94.80} \\ 
		&                           & 40\%       & 92.55        & 90.37 & 73.23 & 92.03 & 91.84 & 92.97 &94.35& \textbf{94.81}\\ 
		&                           & 60\%       & 93.25         & 90.57 & 77.17 &92.26 & 92.01 & 93.30 &94.37& \textbf{94.84}\\ 
		&                           & 80\%       & 93.48         & 90.74 & 78.20 & 92.39 & 92.15 & 93.77 &94.63& \textbf{95.17} \\ 
		\\[-1em]
		\cline{2-11}
		\\[-1em]
		& \multirow{4}{*}{Micro-F1} & 20\%        & 92.14        & 90.53 & 76.98 &92.23 & 92.30 & 93.08 &94.59& \textbf{95.19}\\
		&                           & 40\%        & 93.09        & 90.83 & 79.61 & 92.55 & 92.46 & 93.50 &94.72&\textbf{95.20} \\ 
		&                           & 60\%       & 93.76      & 91.01 & 81.62 & 92.79 & 92.65 & 93.83 &94.75& \textbf{95.23}\\
		&                           & 80\%       & 93.94         & 91.15 & 82.22& 92.92  & 92.78 & 94.27 &94.98& \textbf{95.52}\\ 
		\\[-1em]
		\hline
		\\[-1em]
		\multirow{8}{*}{ACM} & \multirow{4}{*}{Macro-F1} & 20\%       &69.95&70.41&89.59 &89.16&90.01&88.01&90.00&\textbf{92.06}\\ 
		&                           & 40\%       &71.15&70.82&89.77&90.14&90.82&89.42&91.49&\textbf{92.33}\\ 
		&                           & 60\%       &71.74&69.67&89.72&90.71& 91.51 &90.39&92.27&\textbf{92.59}\\ 
		&                           & 80\%       &72.18&67.23&89.42&91.01& 91.71 &90.79&92.70&\textbf{93.08} \\ 
		\\[-1em]
		\cline{2-11}
		\\[-1em]
		& \multirow{4}{*}{Micro-F1} & 20\%       &72.12&74.02&89.47 &89.12&89.89&88.08&90.03&\textbf{91.94} \\
		&                           & 40\%      &73.17&74.57&89.65&90.11& 90.73 &89.48&91.57&\textbf{92.27}\\
		&                           & 60\%      &73.65&74.10&89.60&90.64& 91.37 &90.42&92.32&\textbf{92.52}\\
		&                           & 80\%      &74.14&72.69&89.29&90.93& 91.56 &90.80&92.73&\textbf{92.97}\\ 
		[-1em]\\
		\hline		
	\end{tabular}
	}
\vspace{-0.1in}
\end{table*}

\subsection{Experimental Settings}

\subsubsection{Compared Methods}
We compare our SHGNN with following homogeneous and heterogeneous graph embedding methods.
\begin{itemize}[leftmargin=1.5em]
	\item \textbf{metapath2vec}~\cite{Dong:2017:MSR:3097983.3098036}: A heterogeneous graph embedding method that performs a random walk on meta-paths and utilizes skip-gram to generate embedding. This model relies on a single user-specified meta-path, so we test all meta-paths separately and report the best results.
\item \textbf{GCN}~\cite{DBLP:conf/iclr/KipfW17}: A homogeneous graph convolutional network (GCN) performs on the meta-path connected homogeneous graphs generated by different meta-paths. We test GCN on different meta-paths and report the results from the best meta-path.
\item \textbf{GAT}~\cite{DBLP:conf/iclr/VelickovicCCRLB18}: A homogeneous GNN, which utilizes the masked attention mechanism on the homogeneous graphs. Similarly, we test GAT on meta-path based homogeneous graphs and report the results from the best meta-path.
\item \textbf{HetGNN}~\cite{zhang2019heterogeneous}: A heterogeneous GNN, which jointly considers heterogeneous node contents encoding, type-based neighbors aggregation, and heterogeneous types combination.
\item \textbf{HAN}~\cite{Wang:2019:HGA:3308558.3313562}: A heterogeneous GNN that learns meta-path specific node embeddings from different meta-paths first and then leverages the attention mechanism to combine them into one vector representation for each node.
\item \textbf{MAGNN}~\cite{fu2020magnn}: A heterogeneous GNN that boosts the graph embedding performance by encapsulating input node attributes, incorporating intermediate semantic nodes, and combining messages from multiple meta-paths.
\item \textbf{MAGNN-AC}~\cite{jin2021heterogeneous}: A state-of-the-art heterogeneous GNN, which uses topological relationship between nodes as guidance to complete attributes for no-attribute nodes via attention mechanism. It can solve missing attributes problem in heterogeneous graphs.
\end{itemize}

\subsubsection{Implementation Details}
We implement our framework using the PyTorch library\footnote{Our source code and datasets are available at: \url{https://github.com/Wentao-Xu/SHGNN}.}~\cite{paszke2019pytorch}, and run on all experiments on a single NVIDIA Tesla V100 GPU.
To eliminate the fluctuations caused by different initialization, we repeat the training and testing procedure 10 times for all the experimental results and report the average values.
For the hyper-parameters, we mainly tune the learning rate and the hyper-parameter $d_1$, which is the dimension of many vectors in the SHGNN. The best learning rate is 0.005; the best $d_1$ is 512 for IMDB, 128 for DBLP and ACM.

\subsection{Node Classification Task (RQ1)}
\label{sec:node_class}
We first compare the performance of different methods on the node classification task, which is a frequently-used task to evaluate the heterogeneous graph embeddings.
We feed the embeddings of labeled nodes (movies in IMDB, authors in DBLP, and papers in ACM) learned by each embedding model to a linear support vector machine (SVM) classifier. 
The proportions of nodes used for training the SVM are range from 20\% to 80\%.
Following the setting of previous work~\cite{Wang:2019:HGA:3308558.3313562,fu2020magnn,jin2021heterogeneous}, only the nodes in the testing set are fed to the linear SVM because the labels of nodes in the training and validation sets have participated in the model training process.

Table~\ref{tab:node_class} shows the results of Macro-F1 and Micro-F1 on IMDB, DBLP and ACM datasets.
We reproduce the results of MAGNN\footnote{\url{https://github.com/cynricfu/MAGNN}} and MAGNN-AC\footnote{\url{https://github.com/liangchundong/HGNN-AC}} with their open-source codes, and we take the results of other baselines from~\cite{jin2021heterogeneous}.
From Table~\ref{tab:node_class} we can find that our SHGNN outperforms other baselines across all training proportions and datasets.
On the IMDB dataset, our SHGNN gains the 3\% to 5\% improvements on Macro-F1 and Micro-F1 compare with the current state-of-the-art MAGNN-AC.
On DBLP and ACM datasets, even the existing methods have obtained high scores over 90\%; our SHGNN still achieves better performance.
The improvements on the IMDB, DBLP and ACM datasets verify that incorporating the graph structural information in the node's feature and the information aggregation on meta-paths can boost the performance of heterogeneous graph embeddings.

\begin{table}[t]
	\caption{The results (\%) of node clustering.}
	\vspace{-0.1in}
	\label{tab:node_clust}
	\resizebox{0.48\textwidth}{!}{
		\begin{tabular}{l|cc|cc|cc}
			\hline
			\\[-1em]
			Dataset& \multicolumn{2}{c|}{IMDB}  &  \multicolumn{2}{c|}{DBLP} &  \multicolumn{2}{c}{ACM}\\
			
			Metrics& NMI& ARI & NMI& ARI & NMI& ARI\\
			\\[-1em]
			\hline
			\\[-1em]
			metapath2vec &	0.89 &0.22 & 74.18&  78.11&-&-\\
			GCM& 7.46 & 7.69 & 73.45 & 77.50 &-&-\\
			GAT& 7.84 & 8.87& 70.73& 76.04 &-&-\\
			HAN& 10.79& 11.11& 77.49 & 82.95&-&-\\
			MAGNN& 14.03& 13.25& 79.63 & 84.58&66.48&71.32\\
			MAGNN-AC&13.82&13.78&80.29&85.47&67.52&72.06\\
			\\[-1em]
			\hline
			\\[-1em]
			\textbf{SHGNN} & \textbf{14.75} & \textbf{16.55}& \textbf{82.50}& \textbf{86.83}&\textbf{72.94}&\textbf{77.41}\\
			[-1em]\\
			\hline
		\end{tabular}
	}
	\vspace{-0.1in}
\end{table}

\subsection{Node Clustering Task (RQ2)}
We further evaluate the performance of our SHGNN on another widely-used task: the node clustering task, to evaluate the heterogeneous graph embeddings.
We feed the embeddings of labeled nodes (movies in IMDB, authors in DBLP, and papers in ACM) generated by each method to the K-Means algorithm. 
We set the number of clusters in K-Means as the number of each dataset's classes (3 for IMDB and ACM, and 4 for DBLP). 
Following the previous work~\cite{Wang:2019:HGA:3308558.3313562,fu2020magnn}, we adopt the normalized mutual information (NMI) and adjusted rand index (ARI) as evaluation metrics.

Table~\ref{tab:node_clust} shows the results of NMI and ARI on IMDB, DBLP, and ACM datasets.
We reproduce the MAGNN and MAGNN-AC with their open-source codes, and we take the results of other compared methods from~\cite{fu2020magnn}.
From Table~\ref{tab:node_clust}, we can see that SHGNN is consistently better than other baselines in node clustering.
Note that all models have much poorer performance on IMDB than on DBLP and ACM. This is probably because every movie node in the original IMDB database has multiple genres, but we only select the first one as its class label. 
Our SHGNN  achieves a 5\% relative improvement on NMI and 20\% on ARI compared with the current best results on the IMDB dataset.
On the ACM dataset, our SHGNN obtains a nearly 8\% relative improvement on NMI and 7\% on ARI compared with the state-of-the-art MAGNN-AC model.
These improvements achieved by SHGNN on the node clustering task further demonstrate the effectiveness of our method.

\subsection{Ablation Study (RQ3)}
To study the effect of some key designs in our framework, we apply an ablation study on our SHGNN model. To be specific, we study the effect of node centrality encoding in Section~\ref{subsec:centrality} and tree-attention aggregator in Section~\ref{subsec:meta-path}.
We remove some components in SHGNN and observe the new experimental results of SHGNN's variants.
For the variants without node centrality encoding, we directly use the node structural features in Section~\ref{subsec:node_structural} as the node's latent vector.
For the variants without a tree-attention aggregator, we calculate the attention values between two end nodes of the meta-path and directly aggregate information from meta-path based neighbors to the target node with the attention values.
Table~\ref{tab:ablation} shows the results of ablation study, and we have the following observations:
\begin{enumerate}[leftmargin=1.5em]
	\item Adding the node centrality encoding to differentiate the importance of nodes in different structural information can boost the performance of heterogeneous graph embeddings.
	\item Both node centrality indicators $c(v)$ and $c^+(v)$ are important and helpful to embeddings.
	\item The tree-attention aggregator can improve the performance, especially on the DBLP dataset. Therefore, incorporating the graph structural information in information aggregation on meta-paths is effective for heterogeneous graph embeddings.
\end{enumerate}

\begin{table}[t]
	\centering
	\caption{The node classification results  (\%) of ablation study. The \cmark and - indicate having or not having the component. The Macro and Micro are the average Macro-F1 and Micro-F1 of different training proportions (20\%, 40\%, 60\% and 80\%).}
	\vspace{-0.1in}
	\label{tab:ablation}
	\begin{tabular}{c c c|cccc}
		\hline
		\\[-1em]	
		\multicolumn{2}{c}{Centrality} &Tree- &\multicolumn{2}{c}{IMDB} &\multicolumn{2}{c}{DBLP}  \\
		$c(v)$&$c^+(v)$&attention&Macro&Micro &Macro&Micro\\
		\\[-1em]	
		\hline
		\\[-1em]	
		-&-&-&59.99&60.12&91.96&92.53\\
		\cmark&-&-&60.93&61.14&92.49&93.09\\
		\cmark&\cmark&-&61.80&61.95&93.22&93.83\\
		-&-&\cmark&61.75&61.86&94.35&94.80\\
		\cmark&\cmark&\cmark&\textbf{62.49}&\textbf{62.62}&\textbf{94.91}&\textbf{95.29}\\
		[-1em]\\
		\hline
	\end{tabular}
	\vspace{-0.15in}
\end{table}

\begin{figure}[t]
	\centering
	\begin{subfigure}[b]{0.22\textwidth}
		\centering
		\includegraphics[width=\textwidth]{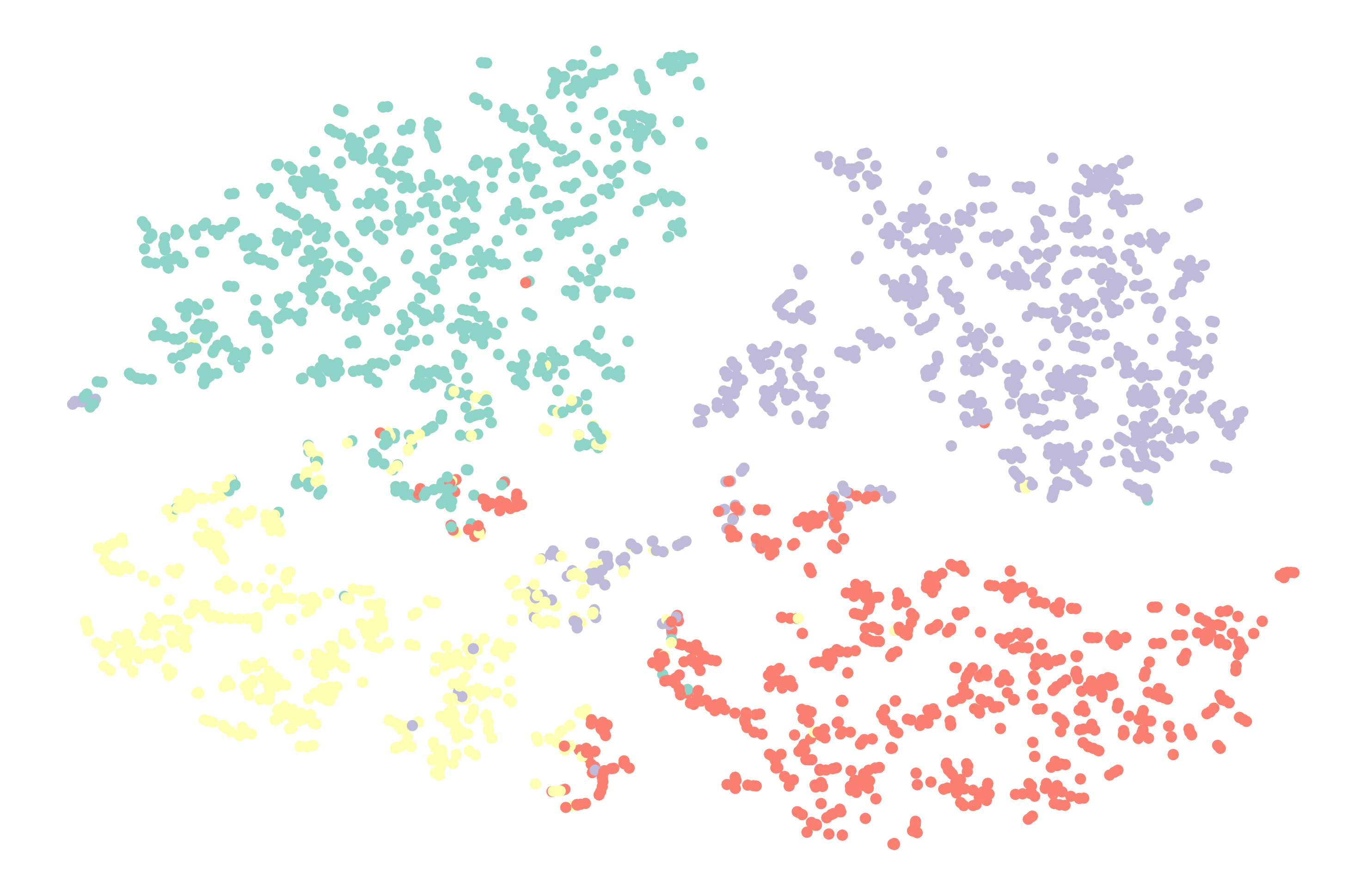}
		\caption{MAGNN}
	\end{subfigure} 
	\hfill
	\begin{subfigure}[b]{0.22\textwidth}
		\centering
		\includegraphics[width=\textwidth]{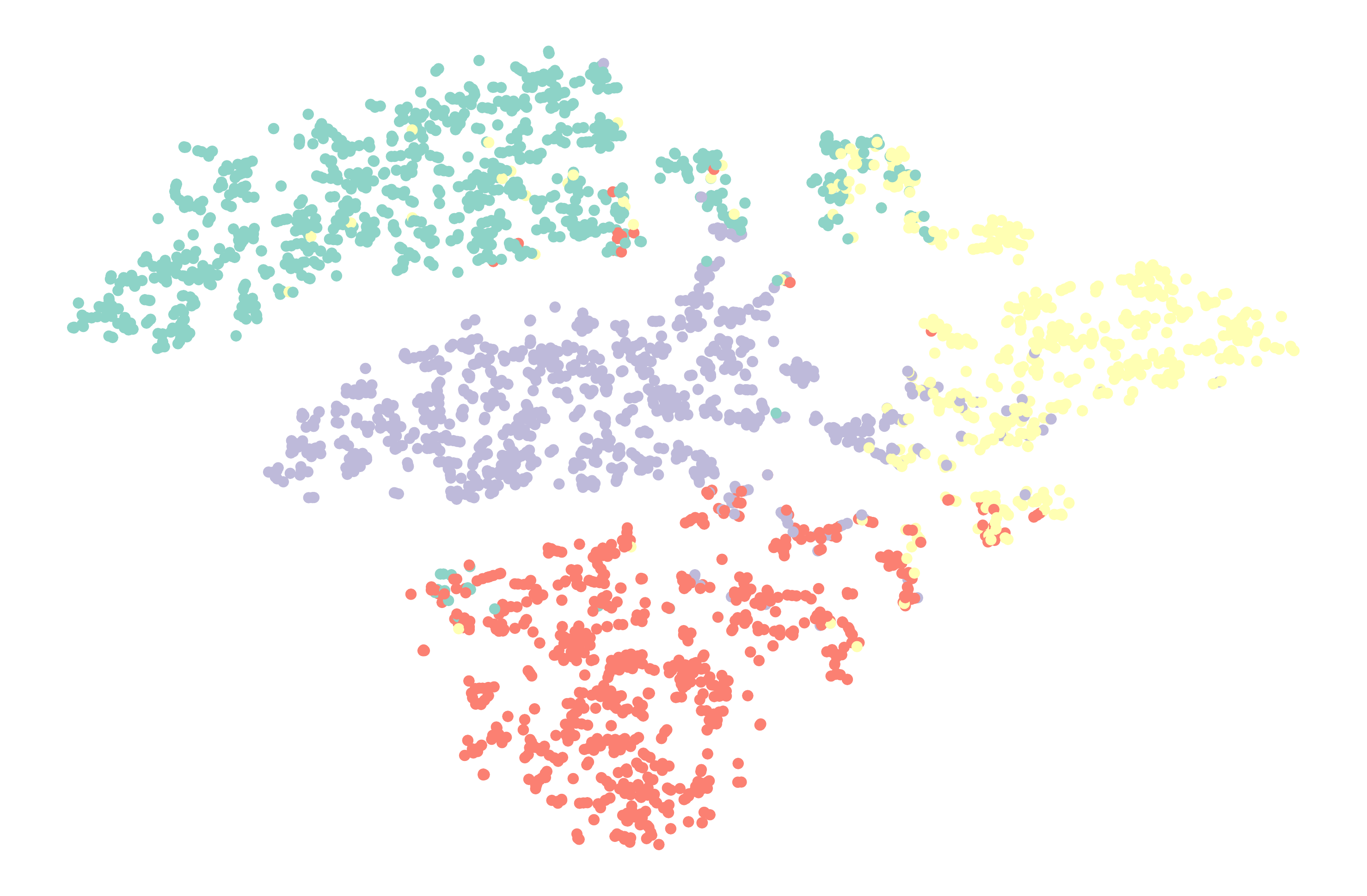}
		\caption{MAGNN-AC}
	\end{subfigure}
	\hfill
		\hfill
\begin{subfigure}[b]{0.22\textwidth}
		\centering
		\includegraphics[width=\textwidth]{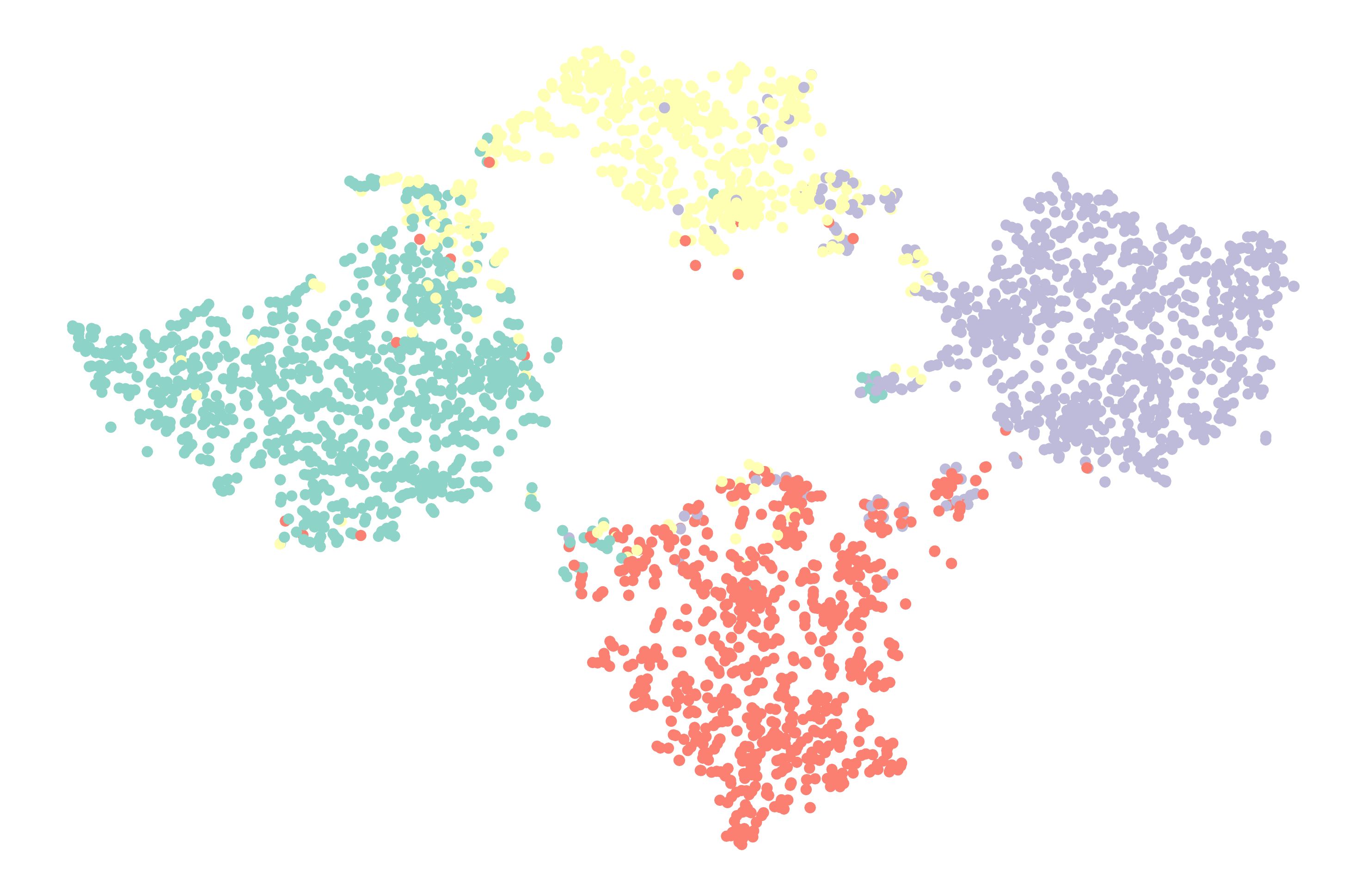}
		\caption{SHGNN}
\end{subfigure}
\vspace{-.1in}
	\caption{Visualization of the Author nodes' embeddings in the test set of DBLP dataset. Each point represents an author and its color indicates the research area of author.}
\vspace{-.15in}
\label{fig:visual}
\end{figure}
\subsection{Visualization (RQ4)}
We visualize the same-dimensional node embeddings of different methods to show a more intuitive comparison. We use the t-SNE~\cite{van2008visualizing} to visualize the embeddings of Author nodes in the test set of the DBLP dataset.
Figure~\ref{fig:visual} shows the results of visualization.
The MAGNN, MAGNN-AC, and SHGNN methods can separate the nodes of different classes, but our SHGNN incorporates the structural information into the node embeddings; thus, our SHGNN can relatively better separate the nodes of various labels.

\section{Conclusion}
We propose a new structure-aware heterogeneous graph neural network (SHGNN) that can incorporate structural information into graph embeddings.
The key insights of our SHGNN include: 1) encoding the node centrality, which measures the importance of nodes; 2) incorporating the graph structure information into the aggregation module on meta-path.
Experimental results of node classification and node clustering tasks on benchmark datasets demonstrate the effectiveness of our SHGNN.
\bibliographystyle{ACM-Reference-Format}
\bibliography{SHGNN}

\end{document}